# Proposition of Augmenting V2X Roadside Unit to Enhance Cooperative Awareness of Heterogeneously Connected Road Users

Keyvan Ansari [1][0000-0002-9969-7682] and Khondokar Fida Hasan [2]

[1] Murdoch University, Murdoch, WA 6150, Australia
`keyvan.ansari@murdoch.edu.au`
[2] Queensland University of Technology, Brisbane, QLD 4000, Australia
`fida.hasan@qut.edu.au`

**Abstract.** Intelligent transportation and autonomous mobility solutions rely on cooperative awareness developed by exchanging proximity and mobility data among road users. To maintain pervasive awareness on roads, all vehicles and vulnerable road users must be identified, either cooperatively, where road users equipped with wireless capabilities of Vehicle-to-Everything (V2X) radios can communicate with one another, or passively, where users without V2X capabilities are detected by means other than V2X communications. This necessitates the establishment of a communications channel among all V2X-enabled road users, regardless of whether their underlying V2X technology is compatible or not. At the same time, for cooperative awareness to realize its full potential, non-V2X-enabled road users must also be communicated with where possible or, leastwise, be identified passively. However, the question is whether current V2X technologies can provide such a welcoming heterogeneous road environment for all parties, including varying V2X-enabled and non-V2X-enabled road users? This paper investigates the roles of a propositional concept named Augmenting V2X Roadside Unit (A-RSU) in enabling heterogeneous vehicular networks to support and benefit from pervasive cooperative awareness. To this end, this paper explores the efficacy of A-RSU in establishing pervasive cooperative awareness and investigates the capabilities of the available communication networks using secondary data. The primary findings suggest that A-RSU is a viable solution for accommodating all types of road users regardless of their V2X capabilities.

**Keywords:** Collision Avoidance, Cooperative Awareness, Cooperative ITS, Heterogeneous V2X Communications, Road Safety, Smart Roads

# 1 Introduction

Cooperative Intelligent Transportation Systems (C-ITS) are critical to enabling safer, greener, and self-driven transportation, which can only be realized if pervasive cooperative awareness is achieved. In this context, cooperative awareness refers to road users' perceptions of and reactions to traffic conditions and road rules, which is an important basis for cooperative driving. Automakers have recently started manufacturing communications-enabled vehicles to reduce road tolls and injuries and enable instantaneous C-ITS services of safety and awareness. The major technologies standardized for Vehicle-to-Everything (V2X) communications include IEEE 802.11p-based Dedicated Short-Range Communications (DSRC) and Long-Term Evolution-Vehicle (LTE-V). V2X communications involve different types of links, with specific requirements each, including Vehicle-to-Vehicle (V2V), Vehicle-to-Infrastructure (V2I) and Vehicle-to-Vulnerable-Road-User (V2VRU). These Device-to-Device (D2D) communications links are established between On-Board Units (OBU) and Roadside Units (RSU) or Hand-Held Units (HHU). Using an underlying V2X technology, Basic Safety Messages (BSM) containing vehicle information, such as speed, position and heading, are exchanged amongst OBUs, RSUs and HHUs.

C-ITS is only as effective as the extent of V2X links established between vehicles, infrastructure, and Vulnerable Road Users (VRU); that is, the efficacy of C-ITS degrades on account of every unconnected vehicle or VRU. In other words, the efficacy of C- ITS is directly proportional to the rate at which V2X-enabled vehicles, or Connected Vehicles (CV), penetrate the fleet. In partial-connected settings where both connected and unconnected vehicles and VRUs are present on roads, CVs do not have access to full C-ITS services that rely on pervasive cooperative awareness. In certain use cases, like intersections and crossings, unborn BSMs of unconnected vehicles or VRUs undermine C-ITS services of safety and awareness and create uncertainty for CVs. This situation is exacerbated by the lack of required infrastructure even if the growth in the CV penetration rate increases.

Another barrier to a sector-wide timely deployment of C-ITS is the standardization of two complementing but incompatible V2X technologies, namely, DSRC (IEEE 802.11p) and LTE-V (LTE PC5 Sidelink). These technologies are considered incompatible because current DSRC and LTE-V radios are essentially unable to exchange BSMs directly with radios of the other V2X technology. However, many countries, including Australia, endorse V2X technology neutrality in the ITS band for efficient use of the spectrum. Considering the current state of the two V2X technology families, neither supports co-channel coexistence, that is, "the ability of devices of a V2X technology to detect and defer to transmissions by devices of other V2X technologies to avoid collisions, and vice versa" [1]. Until the futuristic V2X radio technologies [2], such as IEEE 802.11bd DSRC and 5G NR C-V2X, are equipped with at least co-channel coexistence strategies, they can operate in only adjacent channels meaning no direct link can be established between DSRC radios and LTE-V radios.

This paper aims to study an intermediate solution to timely deployment of C-ITS where DSRC and LTE-V compete on neutral ground, which involves an all-rounder communications bridge, e.g., one per each smart road segment, which decodes and encodes all different types of V2X messages. This study considers only Mode 4 of LTE PC5 Sidelink, which defines the standalone LTE PC5 air interface for direct communications between vehicles without requiring them to connect to a local LTE base station. Throughout the paper, we refer to V2X communications using LTE-V PC5 Sidelink Mode 4 as Cellular Vehicle-to-Everything (CV2X).

The remainder of this paper is organized as follows. Section 2 discusses major barriers to large-scale V2X deployments. In Section 3, we investigate an Augmenting RSU (A-RSU) as an empirical solution to the issues of DSRC and CV2X coexistence and early-stage deployment of C-ITS. Section 4 investigates the data duplication and filtering challenges in the utilization of A-RSU. Section 5 examines heterogeneous data links established amongst road users with the aid of A-RSU. Section 6 presents validation results for system latency of data links between indirectly connected vehicles. Finally, section 7 concludes the article.

## 2 Barriers to large-scale V2X deployment

Without a doubt, it is challenging to deploy V2X radios in new and existing vehicles all at once. Thus, a mixed traffic of V2X-enabled and legacy vehicles is unavoidable once V2X RSUs are also deployed at critical and necessary locations. Nevertheless, the rate at which BSMs are received at a connected vehicle, that is, the reception frequency of Cooperative Awareness Messaging (CAM), is crucial for the overall performance of C-ITS. In other words, road users not connected with others to the extent necessary for pervasive cooperative awareness degrade the effectiveness of C-ITS at their initial deployment phase. For this, C-ITS cannot reach their full potential of safety and awareness applications while non-V2X-enabled road users are excluded from CAM. In addition to the penetration rate issue, the interoperability of the V2X technologies is another key issue. The functional definition given for 'interoperability' between DSRC and CV2X radios in [1] that is "the ability of devices of a V2X technology to decode at least one mode of transmission by devices of other V2X technologies, and vice versa" is essentially out of reach with the available V2X radio technologies.

Other than the choice or availability of DSRC and CV2X, research and analyses on V2X performance have suggested that neither of these technologies can always individually support C-ITS applications in all traffic scenarios [1, 3]. For instance, DSRC struggles to perform efficiently in dense urban scenarios, while CV2X networks overload quickly under high-frequency messaging scenarios [1]. To support greater efficiency and reliability, car manufacturers may choose to offer a hybrid OBU of DSRC and CV2X to compensate for the disadvantages of both technologies. Although hybrid-OBUs would be considered the optimum solution, their early wide deployment will not be viable due to the higher total cost of ownership than single-technology OBUs. Thus, the need for intercommunication between DSRC and CV2X radios of different road users still exists, which will depend on an efficient interoperability mechanism between the two technologies. Preliminary research on providing interoperability between DSRC and CV2X at the PHY layer is underway, which is limited in scope and applicability. The solution proposed in [4], for example, is designed for stationary network devices unsuitable for C-ITS use cases. Research efforts in this area, however, have resulted in the design of various architectures supporting the coexistence between similar communications technologies, such as Wi-Fi and cellular [5-7].

A common approach investigated in the literature and practice includes vision-enabled or sensing-enabled roadside infrastructure to generate CAM for non-V2X-enabled road users. To do so, such solutions utilize deep/machine learning to detect and classify non-connected road users and derive their position, speed and heading trajectories for the integrated RSU to communicate with connected road users. However, solutions such as Proxy CAM [8], intelligent roadside unit [9], AutoC2X [10] and roadside perception unit [11] suffer from a limiting capability. Although the cooperative intersections and road segments technology, such as the smart intersection at the Texas A&M University RELLIS Campus [12] and the Ann Arbor Smart Intersections Project [13], is emergent in creating smart driving environments, the existing systems lack a holistic approach to road user inclusion. One missing aspect of these solutions is the inward communications link to non-V2X-enabled objects from the C-ITS infrastructure. In other words, nonnative-V2X vehicles, i.e., not equipped with V2X radios, are excluded from CAM in those solutions. A viable approach to establishing the missing communications link between the infrastructure and nonnative- V2X objects is to utilize ubiquitously existing smartphones carried by road users.

The concept of A-RSU is investigated in the following section. In the interest of simplicity, we will refer to all different types of road users, including cars, trucks, bikes, bicycles, pedestrians, etc., as 'road users' throughout the rest of the paper unless it is necessary to make a distinction between the types. In terms of V2X connectivity, road users can be categorized as follows from the perspective of a cooperative intersection/smart road segment system: (1) native-V2X connected road users that are DSRC-enabled and/or CV2X-enabled, (2) nonnative-V2X connected road users who are connected to the cooperative awareness platform via cellular links or others like Bluetooth, and (3) non-connected road users who can only be detected passively by cameras or proximity sensors. In the case of non-connected road users, although there are different ways of detecting and tracking non-connected road traffic, such as the techniques summarized in [14], they are excluded from receiving intercommunications individually. That is safety messages can be communicated to non-connected road users only via electronic road signs, which is not practical at all locations. Nevertheless, considering the rate of cellular connectivity amongst road users in urban areas, the percentage of non-connected road users is negligible, particularly in developed/developing countries.

## 3 Augmenting V2X roadside unit

This paper investigates an Augmenting V2X Roadside Unit (A-RSU) as a measure to address the issues caused by the lack of compatibility and interoperability between V2X communications technologies and the effect of both nonnative-V2X connected and non-connected road users. Such an augmenting system benefits not only the deployment phase of native-V2X connected road users but also nonnative-V2X connected road users and contributes to road safety in the autonomous driving era. Most C-ITS safety applications require a CAM latency of less than either 0.1 seconds or 1 second [15]. For instance, the required CAM latency for time-critical applications, such as 'Emergency Electronic Brake Light' (EEBL), 'Forward Collision Warning' (FCW) and 'Intersection Movement Assist' (IMA) is at most 0.1 sec and for time-sensitive applications, such as 'Blind Spot Warning' (BSW), 'Lane Change Warning' (LCW) and 'Do Not Pass Warning' (DNPW) is at most 1 sec. The aim of A-RSU is to achieve an acceptable low latency level in exchanging road safety data amongst heterogeneous-V2X-enabled and non-V2X-enabled road users. A-RSU is essentially an infrastructure-based extension component of the Cloud Computing on Cooperative Cars (C4S) platform proposed in [16] that enables V2X local fogs and Internet-based cloud services. Enabling cooperative awareness, however, requires relatively high bandwidth, which becomes challenging to meet in congested scenarios at local fogs due to its demand.

Figure 1 depicts a C-ITS setting with heterogeneous communications links, where a group of road users is equipped with DSRC, another group is equipped with CV2X and road users with cellular connectivity exist. A-RSU has DSRC and CV2X radios, connects to the cellular network, and is equipped with surveillance cameras to detect non-connected road users. For the purpose of discussions in this article, we assume A-RSU performs accurate detection of road users' positioning, speed and heading and precisely classifies detected road users using an Image Processing Unit (IPU), the details of which is beyond the scope of this article. It is, however, worthwhile to note that although SAE J3224 'V2X sensor-sharing for cooperative & automated driving' [17] is a work in progress at the time of writing, some initial works on 'cooperative awareness' [18-22] proposed mechanisms to directly share raw sensory data (of non-V2X-enabled road users) among native-V2X connected road users. Interested readers are referred to [9, 11], where comprehensive reviews on cooperative awareness concepts are provided.

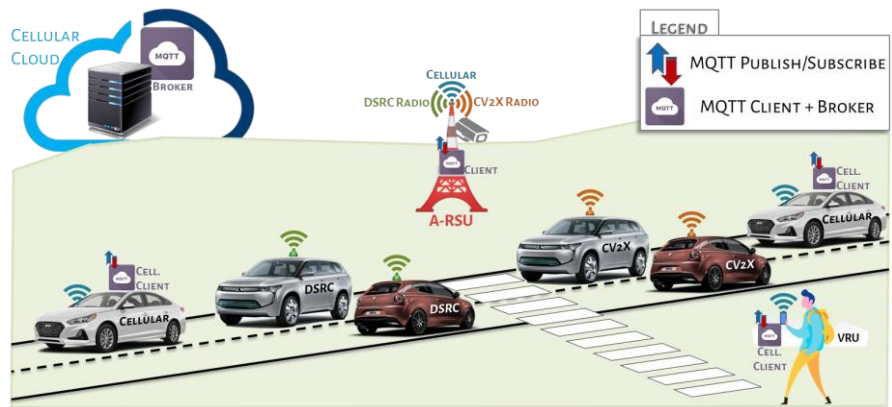

**Fig. 1.** Design solution for early C-ITS deployment of heterogeneous V2X and nonnative-V2X technologies

### 3.1 Design essentials

Although the Society of Automotive Engineers (SAE) has released several standards to specify different requirements for V2X communications, the SAE J2945/1 standard 'On-Board System Requirements for V2V Safety Communications' determines the Maximum Inter-Transmission Time interval (MAX-ITT) to be 600 ms [23]. The MAX-ITT is determined based on the information update requirements of time-sensitive safety-critical V2X applications. Therefore, it is essential for the A-RSU system to keep the overall end-to-end V2X communications delay below 600 ms for all safety applications. The overall end-to-end latency and communications link latency should include the overhead for both the camera system and the detection/processing functions of the IPU. When detecting road users by camera, the tracking capabilities of the camera system are limited by the performance of the IPU, that is, the number of road users possible to be detected and tracked per second. The performance of the camera system depends on the number of cameras being utilized, and the performance of the IPU, i.e., the tracking time, depends on the number of road users being tracked. This additional computation time is reported to be less than 300 ms for the latency of multi-camera vehicle tracking systems and the IPU's detection/processing functions combined [24].

### 3.2 Implementation challenges

Although the function of a nominal IPU on A-RSU is beyond the scope of this study, we note that, while and if the IPU is reliant on calibrations, such a road user detection system would require an internal monitoring system to ensure the accuracy and integrity of position estimation of road users. In addition, precisely distinguishing and deriving the location coordinates of simultaneously arriving vehicles in multiple parallel lanes is a challenge that would lead to image overlapping with even modern image recognition algorithms based on advanced computer vision and deep-learning methods. While traffic of only V2X-connected road users does not suffer from the two aforementioned issues, as for the second issue, in a fully V2X-connected environment, vehicles are easily differentiated by the Transmitter_ID field of BSM.

To address the road user detection issue for C-ITS deployment in the presence of both connected and non-connected road users, the IPU could identify the latter road user type through filtering out the IPU's estimated positions of road users from the historical data of connected road users. To this end, A-RUS maintains a historical list of timestamped position data of connected road users for a brief period, e.g., 200 ms, during which the position estimates of all IPU detected road users are compared with the historical data. The position estimates for potential non-connected road users, i.e., not conformable to the historical data, are further compared against the data received over the next few tens of milliseconds, e.g., 100 ms, before a road user can be tagged as non-connected to allow for late arrival of BSMs from connected road users. If the position estimates of road user X ($RUX^{IPU}$) conform to a piece of historical data for connected road user Y ($RUY^{BSM}$), i.e., the difference between $RUX^{IPU}$ and $RUY^{BSM}$ be less than the calibration error, then road users X and Y are the same and hence connected; otherwise, $RUX^{IPU}$ is a non-connected road user.

## 4 The inconsistent position accuracy problem

Considering A-RSU is equipped with various communications and road user recognition technologies, it is possible to duplicate BSMs that identify the same road users with inconsistent position, speed and heading trajectory data are disseminated if A-RSU attempts to generate BSMs on behalf of connected road users. To avoid the problem, A-RSU should not generate BSMs on behalf of connected road users, i.e., it only retransmits (translates and relays) BSMs received from both native and nonnative-V2X connected road users in the vicinity to augment situational awareness of all road users. This restriction on generating BSMs on behalf of connected vehicles is essential to avoid confusion due to receiving inconsistent duplicate BSMs for one single road user. A-RSU may generate BSMs only on behalf of non-connected road users.

The SAE J2735 standard defines fifteen types of messages for inter-vehicular communications, among which BSM is designated for V2X safety applications. BSM enables safety applications by communicating critical information about vehicles, such as current vehicle state data, including 'latitude', 'longitude', 'elevation' and 'position accuracy' as well as 'heading' [25]. The 'position accuracy' field includes quality parameters to derive positioning accuracy for each axis. For instance, Dilution of Precision (DOP), that is of two types of positional (suitable for standalone positioning) and time (suitable for relative positioning), is a unitless indicator of the quality of position measurements based on satellite constellation geometry where higher values yield poorer positioning solutions. The IPU positioning error ($\sigma$), aka calibration error, depicted in Figure 2, is the difference between the actual point positioning, i.e., by means of a global navigation satellite system (GNSS), which is included in BSMs and the point position calculations by the IPU.

The actual point position perimeter shown in Figure 2 naturally contains positional errors (horizontal and vertical) due to GNSS measurement errors. Although the accuracy of GNSS-based positioning systems depends on several factors, terrestrial receivers are typically accurate to within a 5 m radius under the open sky, while real-time positioning within a few centimetres is possible using dual-frequency receivers and/or augmentation systems. However, both GNSS positional errors (for connected vehicles) and IPU positioning errors (for non-connected vehicles), if beyond a predicted level, could become the source of inconsistencies in identifying non-connected road users precisely. Such inconsistencies would mislead the A-RSU to assume a connected vehicle is not connected, which leads to the creation of a ghost road user (see Figure 2). Addressing the ghost road user problem is an open issue and beyond the scope of this study.

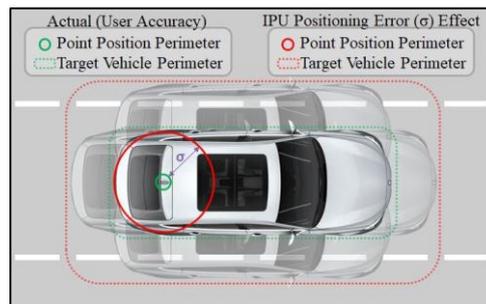

**Fig. 2.** IPU positioning error effect

## 5 A-RSU support of heterogeneous links

As sketched in Figure 1, there exist various peer-to-peer (homogenous) links and infrastructure-based links involving A-RSU (heterogeneous) have been established among road users. Of the heterogeneous links, we consider 10 intercommunicating scenarios to evaluate the feasibility of the A-RSU system (presented in Section 6). Table 1 illustrates these 10 possible V2X intercommunicating scenarios involving incompatible V2X technologies in the presence of nonnative-V2X connected and non-connected road users. A-RSU is the internetworking module required to enable communications amongst incompatible V2X technologies, i.e., DSRC and CV2X, as well as other connected and non-connected road users. To enable communications between dissimilar native-V2X and nonnative-V2X connected road users, the internetworking module should involve the cellular cloud, i.e., Message Queue Telemetry Transport (MQTT) machine-to-machine (M2M)-based protocol [26], in addition to the A-RSU's V2X radios. The cellular cloud enables connections between nonnative-V2X road users and A-RSU via MQTT clients and brokers using MQTT's publish/subscribe methods. MQTT is a reliable, lightweight M2M protocol supporting different Quality of Service (QoS) levels.

**Table 1.** Heterogeneous V2X intercommunicating scenarios

|     | *Uplink* | *Internetworking Module* | *Downlink* |
| --- | --- | --- | --- |
| *1.* | DSRC | A-RSU | CV2X |
| *2.* | DSRC | A-RSU -> Cellular Cloud | Cell. |
| *3.* | CV2X | A-RSU | DSRC |
| *4.* | CV2X | A-RSU -> Cellular Cloud | Cell. |
| 5. | Cell. | Cellular Cloud -> A-RSU | CV2X |
| *6.* | Cell. | Cellular Cloud -> A-RSU | DSRC |
| 7. | Cell. | Cellular Cloud | Cell. |
| 8. | Cam | A-RSU | CV2X |
| 9. | Cam | A-RSU | DSRC |
| *10.* | Cam | A-RSU -> Cellular Cloud | Cell. |

Figure 3 presents the MQTT publish-subscribe model for V2X environments involving nonnative-V2X road users. This MQTT-based telemetry data platform is used to transfer positioning data/estimates in a message format from publishers to subscribers via the broker across TCP/IP connections. Both A-RUS and nonnative-V2X connected road users, aka publishers, publish messages containing positioning data with specifying topic names that identify the source of positioning data/estimates to the central broker (server). The MQTT broker then transmits (pushes) the published messages to the subscribers based on their subscription topics. This system specifies four topics, including (1) 'IPU' that is used for the data published by the A-RSU's IPU on behalf of non-connected road users, (2) 'DSRC' that is used for the data published by the A-RSU on behalf of DSRC-enabled road users, (3) 'CV2X' that is used for the data published by the A-RSU on behalf of CV2X-enabled road users, and (4) 'Cell' that is used for the data published by nonnative-V2X connected road users.

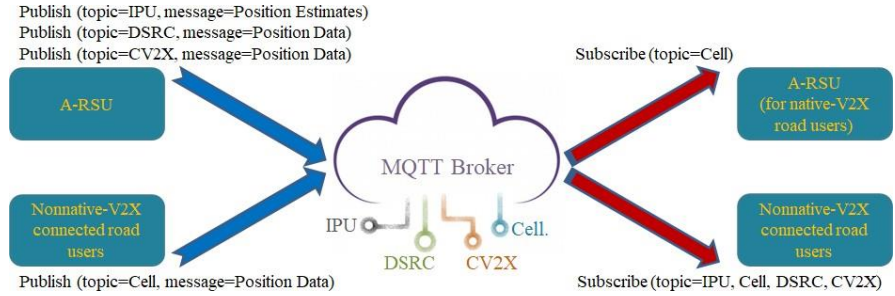

**Fig. 3.** The MQTT publish-subscribe model for heterogeneous V2X environments

Table 2 summarizes the uplink and downlink message types at road objects, including A-RSU, DSRC, and CV2X-connected road users and nonnative-V2X-connected road users. Note that non-connected road users can only be detected passively by the IPU based on which the A-RSU generates BSMs on behalf of non-connected road users. Table 3 contains the A-RSU relaying rules. As discussed, the A-RSU primarily acts as the intercommunicating backbone of smart road segments. That is, the A-RUS translates and relays BSMs disseminated from road users as well as position estimates for non-connected road users.

## 6 Validation of data link latencies

This section investigates the latencies of heterogeneous data links established between road users of different V2X capabilities and examines whether the A-RSU can fulfil the stringent safety requirements of C-ITS safety applications in terms of data latency. We model, i.e., curve fitting, the round-trip time (RTT) indicating link delays for communications technologies in question-based on the measurement results reported in the literature [27-30]. For instance, [27] provides progressive results on average RTT delay for LTE cellular, which we used to model the average RTT delay for 'MQTT over LTE' in combination with the realistic results reported in [28] for MQTT delay. Figure 4 presents the trends/formulas for RTT for each technology separately. Accordingly, Table 4 illustrates link delays for mixed V2X connections assuming both uplink and downlink constitute 50% of each RTT on average.

**Table 2.** Message types for pervasive cooperative awareness

| | A-RSU | Native-V2X connected road user (DSRC) | Native-V2X connected road user (CV2X) | Nonnative-V2X connected road user | Non-connected road user |
|---|---|---|---|---|---|
| Uplink | $BSM_{Tx}^{DSRC}$, $BSM_{Tx}^{CV2X}$, $BSM_{Publish \times 3}^{MQTT}$ | $BSM_{Tx}^{DSRC}$ | $BSM_{Tx}^{CV2X}$ | $BSM_{Publish \times 1}^{MQTT}$ | $Image_{Passive}^{Camera}$ |
| Downlink | $BSM_{Rx}^{DSRC}$, $BSM_{Rx}^{CV2X}$, $BSM_{Subscribe \times 1}^{MQTT}$, $Image_{Passive}^{Camera}$ | $BSM_{Rx}^{DSRC}$ | $BSM_{Rx}^{CV2X}$ | $BSM_{Subscribe \times 4}^{MQTT}$ | N/A |
| representation format: $data\_type_{comm\_link}^{comm\_tech}$ | | | | | |

**Table 3.** A-RSU Relaying Rules

| | Rx | | Tx |
|---|---|---|---|
| 1. | $BSM_{Rx}^{DSRC}$ | → | $BSM_{Tx}^{CV2X}$ + $BSM_{Publish(DSRC)}^{MQTT}$ |
| 2. | $BSM_{Rx}^{CV2X}$ | → | $BSM_{Tx}^{DSRC}$ + $BSM_{Publish(CV2X)}^{MQTT}$ |
| 3. | $BSM_{Subscribe(Cell)}^{MQTT}$ | → | $BSM_{Tx}^{DSRC}$ + $BSM_{Tx}^{CV2X}$ |
| 4. | $Image_{Passive}^{Camera}$ | → | $BSM_{Tx}^{DSRC}$ + $BSM_{Tx}^{CV2X}$ + $BSM_{Publish(IPU)}^{MQTT}$ |

As discussed earlier, C-ITS safety applications are of two types: time-critical and time-sensitive. The key identifier of these C-ITS application categories is their communications latency requirements. Table 4 categorizes the heterogeneous V2X links into 'near real-time' uplink/downlink bundles (i.e., latencies of less than 100 ms) and 'reduced latency' uplink/downlink bundles (i.e., latencies between 100 and 600 ms). The 'near real-time' bundles support both the time-critical applications, including EEBL, FCW and IMA, and the time-sensitive applications, while the 'reduced latency' bundle's support only the time-sensitive applications,

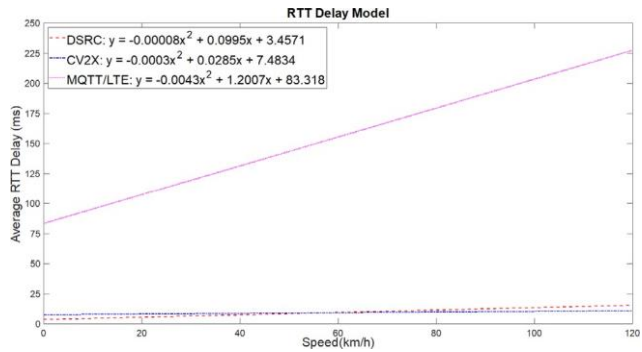

**Fig. 4.** Modeling of average link RTT. Note that the formulas presented here are of general nature for dynamic scenarios and a multitude of affecting factors, such as packet size, transmission frequency, line of sight clearance, and traffic density, among others, are not purposefully considered in the formation of the RTT delay models.

**Table 4.** Link Delay for Heterogeneous V2X

| | Heterogeneous Uplink/Downlink | | Average V2X link delay (ms) | | | | |
|---|---|---|---|---|---|---|---|
| | | | 0 km/h | 30 km/h | 60 km/h | 90 km/h | 120 km/h |
| 1. | DSRC | CV2X | 5.470 | 7.354 | 9.166 | 10.906 | 12.574 |
| 2. | DSRC | Cell.^ | 43.387 | 60.919 | 74.509 | 84.157 | 89.863 |
| 3. | CV2X | Cell.^ | 45.400 | 61.903 | 74.536 | 83.299 | 88.192 |
| 4. | Cell.^ | Cell.^ | 83.318 | 115.469 | 139.880 | 156.551 | 165.482 |
| 5. | Cam* | DSRC | 301.728 | 303.185 | 304.569 | 305.882 | 307.122 |
| 6. | Cam* | CV2X | 303.742 | 304.169 | 304.597 | 305.024 | 305.452 |
| 7. | Cam* | Cell. | 341.659 | 357.735 | 369.940 | 378.275 | 382.741 |

^ LTE technology
* The latency for IPU's detection/processing functions is considered to be 300 ms.

Link Delay Categories:
- < 100ms — near real-time
- < 600ms — reduced latency
- > 600ms — unserviceable

including BSW, LCW and DNPW, for connected road users. Communications solely based on MQTT over LTE (row 4 of Table 4) or involving the IPU (rows 5-7 of Table 4) are not serviceable to time-critical applications and can satisfy only the requirements of time-sensitive applications. However, A-RSU enables connected road users to exploit richer context-aware safety applications. It is important to note that the underlying communications technology considered for MQTT links in this study is LTE and while the solution is viable generally, it is not always serviceable to the C-ITS time-critical applications. However, such a solution (MQTT over cellular links) would become serviceable to more safety applications with the advancements coming with 5G and above, including ultra-low-latency radio access technologies.

## 7 Conclusion

This article investigated the so-called Augmenting V2X Roadside Unit (A-RSU) as a solution to the DSRC/CV2X coexistence issue and the C-ITS early-stage deployment. A-RSU enables connected road users (utilizing either DSRC, CV2X or cellular technologies) to experience a richer context-aware environment. To validate the viability of A-RSU, secondary data, including experimental results of different traffic scenarios reported in the literature, was used to model link latencies for the communications technologies in question. Several C-ITS safety applications were then examined under the properties of A-RSU, and it was concluded that A-RSU is a viable solution and serviceable to both time-critical and time-sensitive applications depending on the bundle of heterogeneous V2X links.

## References

1. Ansari, K., Joint use of DSRC and C-V2X for V2X communications in the 5.9 GHz ITS band. IET Intell Transp Syst., 2021. 15(2): p. 213-224.
2. Naik, G., B. Choudhury, and J.M. Park, IEEE 802.11bd & 5G NR V2X: Evolution of Radio Access Technologies for V2X Communications. IEEE Access, 2019. 7: p. 70169-70184.

3. Cavalcanti, D., et al., Issues in integrating cellular networks WLANs, AND MANETs: a futuristic heterogeneous wireless network. IEEE Wireless Communications, 2005. 12(3): p. 30-41.
4. King, H., K. Nolan, and M. Kelly. Interoperability Between DSRC and LTE for VANETs. in 2018 IEEE 13th International Symposium on Industrial Embedded Systems (SIES). 2018.
5. Almeida, E., et al. Enabling LTE/WiFi coexistence by LTE blank subframe allocation. in 2013 IEEE International Conference on Communications (ICC). 2013.
6. Kaur, H. and J. Malhotra. Coexistence issues and challenges amongst cellular, WiFi and WiMAX networks. in 2014 International Conference on Advances in Engineering & Technology Research (ICAETR - 2014). 2014.
7. Sagari, S., I. Seskar, and D. Raychaudhuri. Modeling the coexistence of LTE and WiFi heterogeneous networks in dense deployment scenarios. in 2015 IEEE International Conference on Communication Workshop (ICCW). 2015.
8. Kitazato, T., et al. Proxy cooperative awareness message: an infrastructure-assisted V2V messaging. in 2016 Ninth International Conference on Mobile Computing and Ubiquitous Networking (ICMU). 2016.
9. Shan, M., et al., Demonstrations of Cooperative Perception: Safety and Robustness in Connected and Automated Vehicle Operations. Sensors, 2021. 21(1): p. 200.
10. Tsukada, M., et al. AutoC2X: Open-source software to realize V2X cooperative perception among autonomous vehicles. in 2020 IEEE 92nd Vehicular Technology Conference (VTC2020-Fall). 2020.
11. Tsukada, M., et al., Networked Roadside Perception Units for Autonomous Driving. Sensors, 2020. 20(18): p. 5320.
12. Smart Intersection. [cited 2021 6 December]; Available from: https://tti.tamu.edu/facilities/smart-intersection/.
13. Lynch, J. $9.95M for 'smart intersections' across Ann Arbor. 2021; Available from: https://record.umich.edu/articles/9-95m-for-smart-intersections-across-ann-arbor/.
14. Sivaraman, S. and M.M. Trivedi, Looking at Vehicles on the Road: A Survey of Vision-Based Vehicle Detection, Tracking, and Behavior Analysis. IEEE Transactions on Intelligent Transportation Systems, 2013. 14(4): p. 1773-1795.
15. Ansari, K., et al. Requirements and Complexity Analysis of Cross-Layer Design Optimization for Adaptive Inter-vehicle DSRC. in Mobile, Secure, and Programmable Networking. 2017. Cham: Springer International Publishing.
16. Ansari, K. Cloud Computing on Cooperative Cars (C4S): An Architecture to Support Navigation-as-a-Service. in 2018 IEEE 11th International Conference on Cloud Computing (CLOUD). 2018.
17. SAE, J3224 V2X Sensor-Sharing for Cooperative & Automated Driving. 2019, SAE International.
18. Li, H. and F. Nashashibi. Multi-vehicle cooperative perception and augmented reality for driver assistance: A possibility to 'see' through front vehicle. in 2011 14th International IEEE Conference on Intelligent Transportation Systems (ITSC). 2011.
19. Al Asif, M. R., Hasan, K. F., Islam, M. Z., & Khondoker, R. (2021, December). STRIDE-based cyber security threat modeling for IoT-enabled precision agriculture systems. In 2021 3rd International Conference on Sustainable Technologies for Industry 4.0 (STI) (pp. 1-6). IEEE.
20. Kim, S.W., et al. Cooperative perception for autonomous vehicle control on the road: Motivation and experimental results. in 2013 IEEE/RSJ International Conference on Intelligent Robots and Systems. 2013.
21. Kim, S., et al., Multivehicle Cooperative Driving Using Cooperative Perception: Design and Experimental Validation. IEEE Transactions on Intelligent Transportation Systems, 2015. 16(2): p. 663-680.

22. Shan, M., S. Worrall, and E. Nebot. Nonparametric cooperative tracking in mobile Ad-Hoc networks. in 2014 IEEE International Conference on Robotics and Automation (ICRA). 2014.
23. SAE, J2945/1 On-Board System Requirements for V2V Safety Communications (revised). 2020, SAE International.
24. Nikodem, M., et al., Multi-Camera Vehicle Tracking Using Edge Computing and Low-Power Communication. Sensors, 2020. 20(11): p. 3334.
25. Ansari, K., C. Wang, and Y. Feng. Exploring dependencies of 5.9 GHz DSRC throughput and reliability on safety applications. in Proceedings of the 10th IEEE Vehicular Technology Society Asia Pacific Wireless Communications Symposium. 2013. IEEE.
26. Ansari, K., Cooperative Position Prediction: Beyond Vehicle-to-Vehicle Relative Positioning. IEEE Transactions on Intelligent Transportation Systems, 2020. 21(3): p. 1121-1130.
27. Xu, Z., et al., DSRC versus 4G-LTE for Connected Vehicle Applications: A Study on Field Experiments of Vehicular Communication Performance. Journal of Advanced Transportation, 2017. 2017: p. 2750452.
28. Sonklin, K., Studies of communication and positioning performance of connected vehicles for safety applications. 2020, Queensland University of Technology.
29. Fan, Y., et al., Network Performance Test and Analysis of LTE-V2X in Industrial Park Scenario. Wireless Communications and Mobile Computing, 2020. 2020: p. 8849610.
30. Shimizu, T., et al., Comparison of DSRC and LTE-V2X PC5 Mode 4 Performance in High Vehicle Density Scenarios, in 26th ITS World Congress. 2019: Singapore.